\documentclass{Interspeech2024}

\interspeechcameraready 

\usepackage{bm}
\usepackage{graphicx}
\usepackage{float}
\usepackage{ascmac} 
\usepackage{fancybox}
\usepackage{float}
\usepackage{tabularx}
\usepackage{booktabs}
\usepackage{amssymb}
\usepackage{multirow}
\usepackage{cite}
\usepackage{multirow}
\usepackage{subcaption}
\usepackage[table,xcdraw]{xcolor}
\usepackage[normalem]{ulem}
\useunder{\uline}{\ul}{}
\usepackage{url}

\def\inR{\in \mathbb{R}}
\def\inRT{\in \mathbb{R}^{T}}
\def\inRTc{\in \mathbb{R}^{T_{c}}}
\def\inRNT{\in \mathbb{R}^{N \times T'}}

\def\inCNN{\in \mathbb{C}^{D \times D}}
\def\inCoN{\in \mathbb{C}^{1 \times D}}
\def\inCNo{\in \mathbb{C}^{D \times 1}}
\def\inCoo{\in \mathbb{C}^{1 \times 1}}

\def\y{{\bm y}}
\def\s{{\bm s}}
\def\shat{\hat{{\bm s}}}
\def\I{{\bm i}}
\def\n{{\bm n}}
\def\cs{{\bm c}^{\text S}}
\def\es{{\bm e}^{\text S}}

\def\A{{\bm A}}
\def\B{{\bm B}}
\def\C{{\bm C}}
\def\D{{\bm D}}
\def\Ab{{\overline{{\bm A}}}}
\def\Bb{{\overline{{\bm B}}}}
\def\C{{\bm C}}
\def\D{{\bm D}}

\def\Zenc{{\bm Z}_{\text{enc}}}
\def\Zsep{{\bm Z}_{\text{sep}}^{\text S}}

\def\TSE{{\tt TSE}}
\def\Encoder{{\tt Encoder}}
\def\Separator{{\tt Separator}}
\def\Decoder{{\tt Decoder}}

\def\param{{\bm \Lambda}}
\def\paramenc{{\bm \theta_{\text{enc}}}}
\def\paramsep{{\bm \theta_{\text{sep}}}}
\def\paramdec{{\bm \theta_{\text{dec}}}}
\def\paramspkenc{{\bm \theta_{\text{spkenc}}}}

\AtBeginDocument{
  \abovedisplayskip     =.4\abovedisplayskip
  \abovedisplayshortskip=.4\abovedisplayshortskip
  \belowdisplayskip     =.4\belowdisplayskip
  \belowdisplayshortskip=.4\belowdisplayshortskip}

\title{SpeakerBeam-SS: Real-time Target Speaker Extraction with Lightweight Conv-TasNet and State Space Modeling}
\name[affiliation={1}]{Hiroshi}{Sato}
\name[affiliation={1}]{Takafumi}{Moriya}
\name[affiliation={1}]{Masato}{Mimura}
\name[affiliation={1}]{Shota}{Horiguchi}
\name[affiliation={1}]{Tsubasa}{Ochiai}
\name[affiliation={1}]{Takanori}{Ashihara}
\name[affiliation={1}]{Atsushi}{Ando}
\name[affiliation={1}]{Kentaro}{Shinayama}
\name[affiliation={1}]{Marc}{Delcroix}

\address{
  $^1$NTT Corporation, Japan}
\email{hrs.sato@ntt.com}


\begin{document}

\maketitle

\begin{abstract}
Real-time target speaker extraction (TSE) is intended to extract the desired speaker's voice from the observed mixture of multiple speakers in a streaming manner.
Implementing real-time TSE is challenging as the computational complexity must be reduced to provide real-time operation.
This work introduces to Conv-TasNet-based TSE a new architecture based on state space modeling (SSM) that has been shown to model long-term dependency effectively.
Owing to SSM, fewer dilated convolutional layers are required to capture temporal dependency in Conv-TasNet, resulting in the reduction of model complexity.
We also enlarge the window length and shift of the convolutional (TasNet) frontend encoder to reduce the computational cost further; the performance decline is compensated by over-parameterization of the frontend encoder.
The proposed method reduces the real-time factor by 78\% from the conventional causal Conv-TasNet-based TSE while matching its performance.

\end{abstract}

\section{Introduction}
While recent advances in machine learning technology have drastically improved speech processing, it is still difficult to handle overlapping speech.
Target speaker extraction (TSE) deals with such overlapping speech. It extracts only the target speaker's voice from the observed mixtures using an enrollment speech to identify the target speaker~\cite{delcroix2018single, wang2018deep, vzmolikova2019speakerbeam, wang2018voicefilter, 10113382}.
TSE is especially promising for personalized applications such as smartphones and smartwatches that are typically required to respond to just their owner.
TSE is also effective in improving teleconferencing by removing background noise and interference speakers from the microphone signals.

Considering real-time speech understanding by machines and telecommunication by humans, it is essential to apply TSE in a causal and streaming manner.
Streaming TSE requires not only low algorithmic latency suitable for streaming applications, but also sufficiently low computational cost at inference time to complete processing in real-time, i.e. real-time factor (RTF) $< 1$.
Due to the general trade-off between computational efficiency and performance, constructing methods that simultaneously maintain both aspects is a challenge.

Despite the difficulty, some preceding works attempted to fulfill the requirements~\cite{wangvoicefilter, eskimez2022personalized, thakker2022fast, dubey2022icassp}.
Wang {\it et al.} first proposed a real-time TSE by devising a lightweight TSE architecture called VoiceFilter-Lite~\cite{wangvoicefilter}.
Since VoiceFilter-Lite enhances filterbank features for ASR, it is not suitable for communication applications.
Subsequently, Eskimez {\it et al.} proposed personalized deep complex convolution recurrent network (pDCCRN)~\cite{eskimez2022personalized}.
pDCCRN efficiently models complex-valued spectrograms by using convolutional neural networks (CNN) to model local dependency and recurrent neural networks (RNN) to model global temporal dependency.
More recently, Liu {\it et al.} proposed the end-to-end enhancement network (E3Net) utilizing a frontend encoder that maps a waveform into a latent representation sequence with 1-d convolutional filters to perform TSE on waveforms~\cite{thakker2022fast}. E3Net has better TSE performance and lower computation cost compared to pDCCRN under conditions when the target speaker is close to the interfering speaker. 
Another causal algorithm that performs TSE in the waveform domain is causal ConvTasNet~\cite{luo2019conv}.
Different from E3Net, which is mainly composed of Long Short-Term Memory (LSTM) layers, the separation network of ConvTasNet consists of multiple blocks of stacked dilated 1-D convolutional layers to attain a sufficiently wide receptive field.
Although Conv-TasNet-based TSE demonstrates superior enhancement performance in causal TSE algorithms compared with VoiceFilter-Lite, pDCCRN, and E3Net~\cite{liu2023quantitative}, it cannot achieve real-time inference on general CPUs. This is due to 1) a high number of 1-D convolutional blocks that must be stacked to ensure a long-enough receptive field, and 2) the short window size and shift of the frontend encoder results in a large number of frames to be processed.

{\it State space modeling (SSM)} has recently been introduced to neural sequence modeling and shown to have superior performance in modeling long-range dependency, compared with traditional deep sequence models such as RNN, CNN, and transformers~\cite{gu2020hippo, gu2021combining, gupta2022diagonal, gu2021efficiently, gu2022parameterization, nguyen2022s4nd}. 
Structured State Space Sequence model (S4)~\cite{gu2021efficiently} and the Diagonal version of S4 (S4D)~\cite{gu2022parameterization}, S4ND~\cite{nguyen2022s4nd} are common SSM variants.
Several speech processing studies use SSM in place of RNNs or CNNs and demonstrated successfully reduced model footprint or increased throughput while maintaining performance, such as in speech recognition~\cite{miyazaki2023structured, shan2023augmenting}, speech generation~\cite{goel2022s}, and denoising~\cite{ku2023multi}.
However, it has yet to be applied to the TSE task.

In this work, we propose a new TSE architecture, SpeakerBeam with State Space modeling (SpeakerBeam-SS); it achieves real-time TSE with high computational efficiency.
By introducing S4D layers to Conv-TasNet-based TSE~\cite{luo2019conv, delcroix2020improving}, fewer dilated convolutional layers are required to capture temporal dependency, resulting in the reduction of model complexity.
To further reduce computational complexity, we adopt a wider window size and shift of the frontend encoder to reduce the number of frames processed, which generally causes a performance decline.
To compensate for this, we adopt over-parameterization of the frontend encoder following~\cite{thakker2022fast}.

An experiment on simulated mixtures of two speakers and noise showed that SpeakerBeam-SS reduced RTF by 78\% compared with the conventional Conv-TasNet-based TSE architecture, while maintaining performance in terms of signal-to-distortion ratio (SDR) and DNSMOS.

\section{Related work}
One preceding work utilizing SSM for speech enhancement is S4ND U-Net~\cite{ku2023multi}. It uses the S4ND~\cite{nguyen2022s4nd} to model multidimensional patterns with SSM. 
The major difference from this study, apart from the task (denoising vs TSE), is that we focus on causal and real-time inference whereas S4ND focuses on reducing the model footprint and not measuring inference speed.
Despite the small footprint, S4ND-UNet was computationally intensive in our experiments; it requires approximately ten times more training time per epoch than the proposed method\footnote{We investigated the extension of S4ND-UNet to TSE tasks by feeding d-vector or speaker vector extracted by ECAPA-TDNN, but training was unstable and 
competitive performance was not achieved.}.

\begin{figure*}[tb]
 \begin{center}
  \includegraphics[width=0.83\hsize]{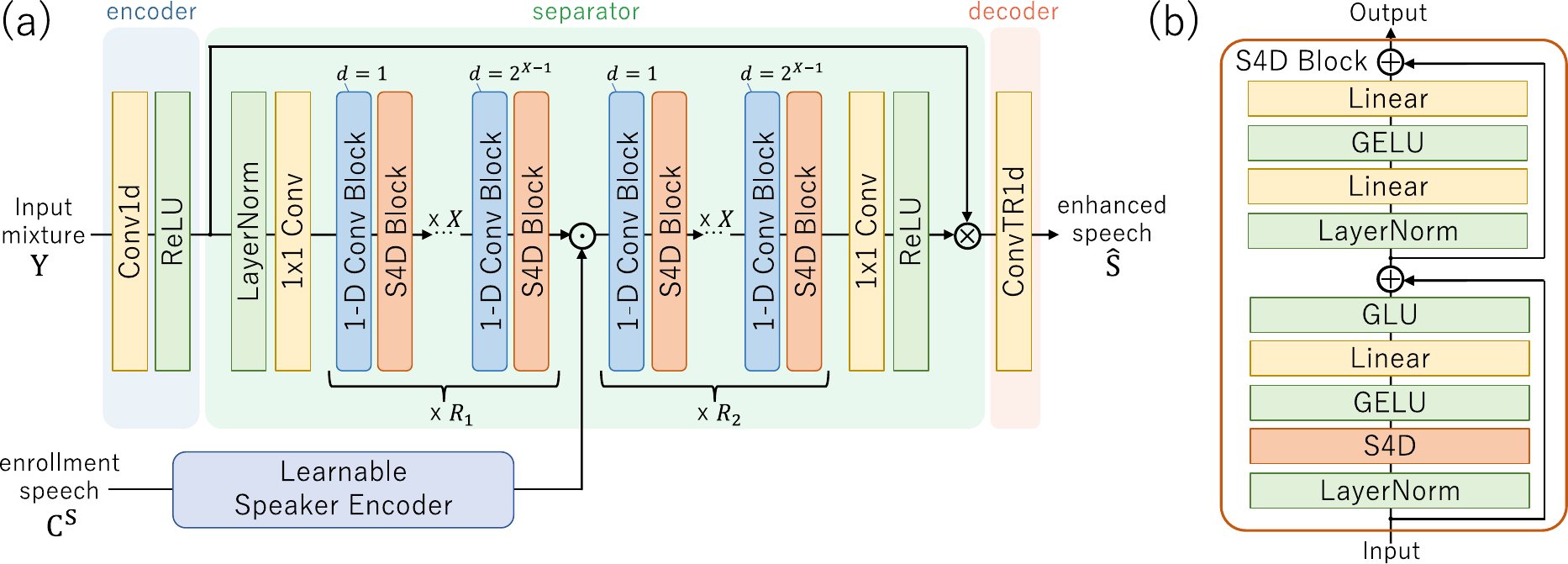}
 \end{center}
 \vspace{-20pt}
 \caption{Overview of the proposed SpeakerBeam-SS architecture. (a) shows the overall structure and (b) shows the details of the S4D block. The dropout layer is omitted from the figure. $d$ refers to the dilation of 1-D convolutional blocks.}
 \vspace{-12pt}
 \label{fig:framework}
\end{figure*}
\section{Background}
\subsection{Conv-TasNet-based TSE}
In this work, we assume that the observed mixture $\y \inRT$ is a single channel signal modeled as $\y = \s + \I + \n$ where $\s, \I, \n \inRT$ are clean target speech, interfering speech, and noise, respectively; $T$ denotes the number of samples in the signal.
TSE aims at extracting only the target speaker's speech from the observed mixture by using the enrollment speech of the target speaker $\cs \inRTc$, as in $\shat = \TSE(\y, \cs; \param)$
where $\TSE(\cdot,\cdot)$ represents the TSE network, $\shat \inRT$ is the enhanced signal, $T_c$ is the number of samples in the enrollment speech, and $\param$ denotes learnable parameters of the model.
In Conv-TasNet-based TSE model~\cite{delcroix2020improving}, the input signal is first projected into a latent space by a frontend encoder consisting of a 1-D convolutional layer as follows:
\begin{align}
    \label{eq:enc}
    \Zenc = \Encoder(\y; \paramenc),
\end{align}
where $\Encoder(\cdot)$ represents the encoder network, $\Zenc \inRNT$ is the latent representation of the input signal, and $\paramenc$ is learnable parameters of the encoder. $N$ is the number of filters in the encoder.
$T'$ is the number of the frames, which is determined by the number of samples in the signal $T$ and the kernel size $L$ and shift $L/2$ of the 1-D convolutional layer.
To identify the target speaker, speaker vector $\es$ is extracted from enrollment speech $\cs$ by a learnable speaker encoder. 
Based on $\es$, the separator network extracts the target speaker as follows:
\begin{align}
    \label{eq:sep}
    \Zsep = \Separator(\Zenc, \es; \paramsep),
\end{align}
where $\Separator(\cdot,\cdot)$ represents the separator network, $\Zsep \inRNT$ is the extracted representation of the target speaker's speech, and $\paramsep$ is the learnable parameters of the separator. 
The separator network in Conv-TasNet consists of multiple stacked dilated 1-D convolutional blocks with increasing dilation from $d=1$ to $d=2^{X-1}$ to ensure a sufficient receptive field where $X$ is the number of repetitions of dilated 1-D convolutional blocks. 
Representation $\Zsep$ is then transformed back into a waveform signal, $\shat$, by a 1-D transposed convolution layer-based decoder as follows:
\begin{align}
    \label{eq:sep}
    \shat = \Decoder(\Zsep; \paramdec),
\end{align}
where $\Decoder(\cdot)$ represents the decoder network, and $\paramdec$ represents the learnable parameters of the decoder.
All parameters $\param = \{\paramenc, \paramsep, \paramdec, \paramspkenc\}$ are jointly trained on the minimization criterion between the enhanced speech $\shat$ and target speech $\s$, where $\paramspkenc$ denotes the learnable parameters of the speaker encoder.

Although Conv-TasNet-based TSE shows prominent performance among TSE networks, its computation cost is too high to attain real-time inferencing on general CPUs.
One cause of this is the high number of frames processed by the separator network; the kernel size $L$ and shift $L/2$ of the frontend encoder are generally as short as 20 and 10 samples, corresponding 1.25 and 0.625 ms at sampling rates of 16 kHz, respectively~\cite{delcroix2020improving}.
Another cause is the large number of repetitions of stacked dilated 1-D convolutional blocks in the separator network needed to attain a sufficient receptive field.

\subsection{State Space Modeling}
S4~\cite{gu2021efficiently} and S4D~\cite{gu2022parameterization} are variants of SSM. 
The start point of SSM is the introduction of ordinary differential equations (ODEs) that map one-dimensional signal $u(t) \longmapsto v(t)$ as follows:
\begin{equation}
\begin{aligned}
    \label{eq:ode}
    {\bm x}'(t) = \A{\bm x}(t) + \B u(t), \hspace{4pt}
    v(t) = \C{\bm x}(t) + \D u(t),
\end{aligned}
\end{equation}
where ${\bm A}\inCNN$, ${\bm B}\inCNo$, ${\bm C}\inCoN$, and ${\bm D}\inCoo$ are state transition matrices that are learnable parameters. ${\bm x}(t) \inCNo$ is the internal state of the state space equation.
The ODEs are discretized by either Bilinear or zero-order hold (ZOH) discretization to permit the use of sequence-to-sequence modeling. The discretized form of Eq.~\eqref{eq:ode} by ZOH discretization is expressed as follows:
\begin{equation}
\begin{aligned}
    \label{eq:ode_disc}
    {\bm x}_k &= \Ab{\bm x}_{k-1} + \Bb u_k, \hspace{4pt}
    v_k = \C{\bm x}_k + \D u_k,
\end{aligned}
\end{equation}
where $\Ab = \exp({\Delta \A}), \Bb = ({\Delta \A})^{-1} (\exp({\Delta \cdot \A})-{\bm I})\cdot{\Delta \B}$ and $\Delta \inR$ is a learnable parameter representing time-step size. According to Eq.\eqref{eq:ode_disc}, S4 and S4D can be calculated causally by a simple linear recurrent neural network at inference, which improves computation efficiency.

By setting initial state ${\bm x}_0$ to ${\bm 0}$, Eq.~\eqref{eq:ode_disc} can be expanded as follows:
\begin{equation}
\begin{aligned}
    \label{eq:ode_exp}
    v_k = \C \Ab^k \Bb u_0 + \C \Ab^{k-1} \Bb u_1 + \dots + \C\Bb u_k + \D u_k.
    \vspace{-3pt}
\end{aligned}
\end{equation}
Although the S4 layer processes the signal recurrently in a sample-by-sample manner at inference time, its training and batch inference can be parallelized by introducing SSM convolutional kernel $\bar{\bm K} := (\C\Bb, \C\overline{\A\B},..., \C\overline{\A}^{K-1}\overline{\B})$ where $\bar{\bm K}$ is the signal length~\cite{gu2021combining}. Output signal ${\bm v} = \{v_0,...,v_{K-1}\}$ can be written in the form of the convolution of the kernel $\bar{\bm K}$ and the input signal ${\bm u} = \{u_0,...,u_{K-1}\}$ as follows:
\begin{equation}
\begin{aligned}
    \label{eq:ode_conv}
    {\bm v} &= {\bm u} * \bar{\bm K} + \D u_k,
\end{aligned}
\end{equation}
It is known that fast Fourier transform (FFT) can calculate Eq.~\eqref{eq:ode_conv} efficiently.
Since the input and the output to the S4 layer are actually the two-dimensional feature, Eq.~\eqref{eq:ode_disc}
and Eq.~\eqref{eq:ode_conv} are independently applied to each feature dimension. 

For S4, state matrix $\A$ is parameterized as a diagonal plus low rank (DPLR) form, which theoretically allows the internal state to memorize the history of the input, resulting in the ability to model long-range dependency.
S4D further simplifies the parameterization of $\A$ to diagonal form, which is also empirically effective in temporal modeling.
Moreover, S4D parameterizes $\B$ as the fixed term $\B = {\bm 1}$.
\section{Proposed method}
\label{sec:proposed}
In this work, we propose a new TSE architecture, SpeakerBeam-SS. 
Figure~\ref{fig:framework} (a) shows an overview. 

SpeakerBeam-SS basically follows the architecture of the Conv-TasNet-based TSE proposed in~\cite{delcroix2020improving}. 
For causal implementation, channel-wise layer normalization and causal convolutional layers are adopted in place of global layer normalization and non-causal convolutional layers.

To reduce computational complexity, we introduce the S4D layer to the Conv-TasNet-based architecture to model long-range dependency more efficiently.
The 1-D convolution blocks of the original Conv-TasNet architecture can only see the context provided by the receptive field of the layers below them. Here we augment the conv-blocks with S4D blocks to allow access to the long context. Consequently, fewer repetitions of blocks are required to model enough length of the temporal context.
The detailed structure of the S4D block is shown in Figure~\ref{fig:framework} (b).
The S4D block structure is based on the S4 block in ~\cite{goel2022s} with the modification of replacing S4 layer with S4D layer.
A preliminary experiment on the Conv-TasNet-based architecture showed that introducing the S4D layer yielded better performance than S4 or DSS~\cite{gupta2022diagonal}.
The design of the 1-D convolutional blocks follows ~\cite{delcroix2020improving}.

To further reduce the computational complexit to meet the requirement of real-time processing, we increased the window size and shift.
Although this incurs a performance penalty, which we compensate by increasing the number of the filters of the frontend encoder as found in \cite{thakker2022fast}. 
This also largely reduces the RTF while maintaining the performance (see Figure~\ref{fig:tradeoff}).

Since the S4D block works in a causal manner, the maximum algorithmic latency of the proposed SpeakerBeam-SS is the same as the window length of the frontend encoder.
We also implemented lookahead for the proposed SpeakerBeam-SS for the applications where lower latency constraints are acceptable, by modifying several 1-D convolutional blocks to be non-causal.


\begin{table*}[tb]
\centering
\sisetup{detect-weight,mode=text}
\renewrobustcmd{\bfseries}{\fontseries{b}\selectfont}
\renewrobustcmd{\boldmath}{}
\newrobustcmd{\BOLD}{\bfseries}
\caption{Performance and model complexity of Conv-TasNet based and SpeakerBeam-SS systems. The results were obtained on mixtures of two speakers and noise. `OP' stands for the over-parameterization of the frontend encoder, corresponding to $N=2048$. $L$ and $N$ represent the window shift and the number of filters in the frontend encoder, respectively. $X$ is the number of repetitions of the blocks in the separator network. 95\% confidence interval (CI) is calculated by the bootstrapping method. Values in the bold letter indicate the best performance within systems with delay of 20 ms or less.}
\vspace{-7pt}
\label{tab:main}
\scalebox{.85}[.85]{%
\begin{tabular}{@{}llcccccrl|S[table-format=2.2]l@{}S[table-format=2.2]@{, }S[table-format=2.2]@{}lS[table-format=1.2]@{\hskip 7pt}*{2}{S[table-format=1.2]}@{}}
\toprule
 &  & \multicolumn{4}{c}{Model architecture} & \multirow{2}{*}{\begin{tabular}[c]{@{}l@{}}Latency\\ 
 $\text{[ms]}$\end{tabular}} & \multicolumn{2}{c|}{Complexity} & \multicolumn{5}{c}{SDR [dB]} & \multicolumn{3}{c}{DNSMOS} \\ \cmidrule(lr){3-6} \cmidrule(lr){8-9} \cmidrule(lr){10-14} \cmidrule(l){15-17} 
 & Method & L & N & X & S4D &  & \#param & RTF & {mean} & \multicolumn{4}{c}{(95\% CI)} & {OVRL} & {SIG} & {BAK} \\ \midrule
 & Mixture & - & - & - & - & - & - & - & -0.45 & \multicolumn{4}{c}{-} & 2.44 & 3.95 & 2.17 \\ \midrule
(a) & Baseline-noncausal & 20 & 256 & 8 &  & {$\infty$} & 8.69 M & - & 13.63 & ( &13.39 & 13.86 & ) & 3.40 & 3.71 & 4.24 \\ \midrule
(b1) & Baseline & 20 & 256 & 8 &  & 1.25 & 8.69 M & 1.67 & 11.10 & ( &10.88 & 11.30 & ) & 2.82 & 3.22 & 3.76 \\
(b2) &  & 320 & 256 & 8 &  & 20 & 8.84 M & 0.40 & 9.86 & ( & 9.64 & 10.07 & ) & 2.76 & 3.14 & 3.82 \\ \midrule
(c1) & Baseline + OP & 320 & 2048 & 8 &  & 20 & 10.91 M & 0.54 & 11.41 & ( & 11.19 & 11.59 & ) & 2.91 & 3.29 & 3.87 \\
(c2) &  & 320 & 2048 & 2 & \textbf{} & 20 & 6.87 M & 0.22 & 9.42 & ( & 9.21 & 9.59 & ) & 2.64 & 3.04 & 3.68 \\ \midrule
(d1) & Baseline + OP + S4D & 320 & 2048 & 2 & \checkmark & 20 & 7.93 M & 0.36 & \BOLD 11.58 & ( & 11.37 & 11.77 & ) & \BOLD 2.95 & \BOLD 3.33 & \BOLD 3.89 \\
(d2) & \hspace{7pt} + lookahead 40 ms & 320 & 2048 & 2 & \checkmark & 60 & 7.93 M & 0.36 & 11.88 & ( & 11.67 & 12.06 & ) & 3.14 & 3.51 & 4.03 \\
(d3) & \hspace{7pt} + lookahead 120 ms & 320 & 2048 & 2 & \checkmark & 140 & 7.93 M & 0.35 & 12.12 & ( & 11.89 & 12.32 & ) & 3.21 & 3.58 & 4.05 \\ \bottomrule
\end{tabular}%
}
\vspace{-8pt}
\end{table*}

\begin{table}[tb]
\centering
\sisetup{detect-weight,mode=text}
\renewrobustcmd{\bfseries}{\fontseries{b}\selectfont}
\renewrobustcmd{\boldmath}{}
\newrobustcmd{\BOLD}{\bfseries}
\vspace{-6pt}
\caption{Performance comparison with existing real-time TSE architectures in SDR [dB] and DNSMOS OVRL values.}
\vspace{-8pt}
\label{tab:comp}
\scalebox{.85}[.85]{%
\begin{tabular}{@{}llcS[table-format=2.2]S[table-format=1.2]S[table-format=2.2]S[table-format=1.2]@{}}
\toprule
 & & \multirow{2}{*}{\begin{tabular}[c]{@{}l@{}}Latency\\ $\text{[ms]}$\end{tabular}\hspace{-3pt}} & \multicolumn{2}{l}{2 speakers + noise} & \multicolumn{2}{l@{}}{1 speaker + noise} \\ \cmidrule(lr){4-5} \cmidrule(l){6-7}
 & Method &  & {SDR} & {OVRL} & {SDR} & {OVRL} \\ \midrule
 & Mixture & - & -0.45 & 2.44 & 3.84 & 2.28 \\ \midrule
(e) & pDCCRN \cite{eskimez2022personalized} & 32 & 10.70 & 2.83 & 10.66 & 2.78 \\
(f) & E3Net \cite{thakker2022fast} & 20 & 9.83 & 2.83 & 10.37 & 2.77 \\
(d1) & Ours & 20 & \BOLD 11.58 & \BOLD 2.95 & \BOLD 11.10 & \BOLD 2.89 \\ \bottomrule
\end{tabular}%
}
\vspace{-8pt}
\end{table}

\section{Experiments}
\subsection{Experimental setup}
\subsubsection{Dataset}
Training and evaluation were performed on simulated mixtures generated by speech recordings from the LibriSpeech corpus~\cite{panayotov2015librispeech} and noise samples from the DNS4 challenge dataset~\cite{reddy2021interspeech}. 
For training, we prepared mixtures of two speakers and noise, where the noise was added at signal-to-noise ratio (SNR) values randomly sampled between 0 and 25 dB.
We evaluated two conditions: 1) target and interference speakers and noise mixtures generated with SNR values between 10 and 20 dB and 2) target speaker and noise mixtures generated at SNR values between 0 and 10 dB.
The signal-to-interference ratio (SIR) between the target and interfering speech was set to randomly sampled values between -5 to 5 dB for training and evaluation data.
We generated 50,000, 3,000, and 2,000 mixtures for training, development, and evaluation datasets, respectively, with no duplicate speakers between each dataset.

\subsubsection{System configuration and training procedure}
For the baseline Conv-TasNet-based TSE and the proposed SpeakerBeam-SS, we set the hyperparameter $B\hspace{-1mm}=\hspace{-1mm}256$, $H\hspace{-1mm}=\hspace{-1mm}512$ and $P\hspace{-1mm}=\hspace{-1mm}3$ following the notation in \cite{luo2019conv}. 
We set $R_1$ and $R_2$ in Figure~\ref{fig:framework} to 3 and 1, respectively. 
We set the number of repetitions of the blocks ${\rm X}$ to 8 in the baseline Conv-TasNet-based TSE and 2 in SpeakerBeam-SS.
We increased the number of filters in the frontend encoder, $N$, from 256 to 4096, and the kernel width, $L$, and shift $L/2$ from 20 and 10 to 320 and 160, respectively.
We implemented SpeakerBeam-SS based on the public Conv-TasNet implementation in~\cite{tasnet} and S4D implementation in~\cite{statespace}.
The learnable speaker encoder consisted of one dilated convolution block and frontend encoder following \cite{delcroix2020improving}.
We implemented 40 ms and 120 ms lookahead for SpeakerBeam-SS by modifying the first or first and second 1-D convolutional blocks to be non-causal.
We set the state space dimension $D$ to 32 and the input dimension of the SSM to 256. The hidden dimension of the fully connected layer in S4D blocks was set to 512.
We used the reduce-on-plateau learning rate scheduler with a peak learning rate of 5e-4.

We evaluated pDCCRN and E3Net for comparison\footnote{It has been reported that E3Net set the restriction that the target speaker should be closer to the microphone than the interference speaker~\cite{thakker2022fast}, yet we examine the TSE performance without this restriction to gain a broader range of insights.}.
We implemented pDCCRN based on the public implementation \cite{DCCRN} by modifying it to receive speaker embeddings. Each CNN's kernel size and number of kernels were set to 5 and $\{16,32,64,128,256,256\}$, respectively. We adopted a 2-layer RNN with 128 units. For STFT, we set the window length to 32 ms and the hop to 16 ms. We used the same learning rate scheduler as mentioned above with peak learning rate of 5e-4.
We implemented E3Net based on the description in \cite{thakker2022fast}. We set the number of filters in the frontend encoder to 2048, kernel width to 320, and shift to 160, respectively. 
We adopted a 4-layer LSTM with 256 units. The hidden dimension of the fully connected block was set to 1024.
We used a cosine annealing scheduler with peak learning rate of 1e-3.
As the speaker encoder of pDCCRN and E3Net, we adopted a publicly available d-vector extractor trained on LibriSpeech~\cite{panayotov2015librispeech} and VoxCeleb 1 and 2~\cite{nagrani17_interspeech, chung18b_interspeech} dataset~\cite{Resemblyzer}\footnote{Preliminary experiments showed that the d-vector extractor performed better than ECAPA-TDNN and the learnable speaker encoder model trained from scratch.}. 
All models were trained with Adam optimizer~\cite{kingma2014adam} for 200 epochs with early stopping if the loss did not decrease within 20 epochs. We evaluated the averaged model against the three models that performed the best on the development set.

\subsubsection{Evaluation details}
As performance metrics, we report SDR\cite{vincent2006performance} and DNSMOS P.835~\cite{reddy2022dnsmos} values. 
Moreover, to precisely measure the practical RTF, we implement the inference algorithm for each system on C++.
We measured the RTF of frame-by-frame inferencing achieved when using one core of an AMD EPYC 7502P based on our C++ implementation, as averaged over 100 audio clips. 
\begin{figure}[tb]
\vspace{0pt}
 \begin{center}
  \includegraphics[width=0.85\hsize]{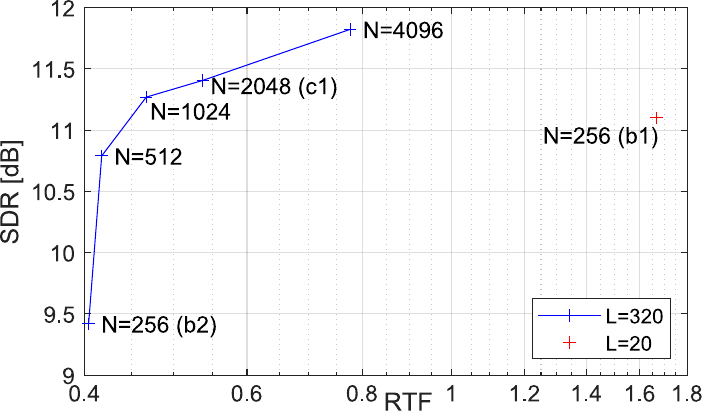}
 \end{center}
 \vspace{-16pt}
 \caption{The relationship between RTF and the enhancement performance with various numbers of filters $N$ and window size $L$, in the frontend encoder.}
 \vspace{-8pt}
 \label{fig:tradeoff}
\end{figure}

\subsection{Experimental results}
Table \ref{tab:main} shows the complexity and the TSE performance for each system. 
Comparing systems (b1) and (b2), enlarging window size $L$ and shift $L/2$ can significantly lower the RTF by reducing the number of frames processed; however, this also results in a significant decrease in performance.
Note that the window size of $L=320$ corresponds to 20 ms of maximum algorithmic delay.
The increase in the number of convolution filters ${\text N}$ from 256 to 2048 (c1) successfully compensates the performance drop while reducing the RTF by about 68\% compared with the baseline system (b1).

The reduction in the repetition number of 1-D convolutional block $X$ from 8 to 2 (c2) lowers the RTF by reducing the computation cost, yet at the same time, the performance significantly drops due to the narrowing of the receptive field.
By introducing S4D blocks, which can efficiently model long sequences, system (d1) attained enhancement performance equal to that of system (c1) with fewer repetitions.
Compared with system (c1), system (d1) achieved about a 27\% reduction in model parameters and a 33\% reduction in RTF while maintaining the enhancement performance.
This result indicates the effectiveness of the S4D layer in constructing real-time TSE.
As a whole, compared to the initial setup of the Conv-TasNet-based TSE models in \cite{delcroix2020improving} (b1),  SpeakerBeam-SS (d1) offers a 78\% reduction in RTF.
Rows (d2) and (d3) show that if lower latency constraints are acceptable, the introduction of lookahead within SpeakerBeam-SS can improve performance.

Table \ref{tab:comp} shows a performance comparison of the proposed SpeakerBeam-SS with other architectures.
In both 2 speaker and noise and 1 speaker and noise conditions, the proposed method outperforms conventional architectures. 
Although E3Net was reported in \cite{thakker2022fast} to have better performance than pDCCRN, this is not the case in our experimental conditions. This is probably because the target speaker is not necessarily closer than the interfering speaker as reported in~\cite{thakker2022fast}.

To further understand the effect of the number of filters in the frontend encoder, we investigated the relationship between RTF and performance with various numbers of filters in frontend encoder $N$ and window size $L$, shown in Figure~\ref{fig:tradeoff}. 
With the conventional setup of the number of filters $N=256$, performance drops substantially when the length of the window is widened from $L=20$ to $L=320$. 
However, the increase in the number of filters $N$ recovers the performance while only modestly impacting the RTF.

\section{Conclusion}
In this work, we investigated reducing the complexity of the Conv-TasNet-based TSE model while maintaining its performance to realize real-time TSE.
The over-parameterization of the frontend encoder while increasing the size of the convolutional kernel and the introduction of state space modeling both successfully reduced RTF without hurting performance. As a result, a 78\% reduction in RTF has been achieved overall compared with the baseline, without hurting the SDR or DNSMOS values.
The proposed SpeakerBeam-SS also shows superior performance to conventional architectures.
The introduction of SSM to other TSE architectures such as pDCCRN and E3Net than ConvTasNet will be part of our future work.

\bibliographystyle{IEEEtran}
\bibliography{mybib}

\begin{thebibliography}{10}
\providecommand{\url}[1]{#1}
\csname url@samestyle\endcsname
\providecommand{\newblock}{\relax}
\providecommand{\bibinfo}[2]{#2}
\providecommand{\BIBentrySTDinterwordspacing}{\spaceskip=0pt\relax}
\providecommand{\BIBentryALTinterwordstretchfactor}{4}
\providecommand{\BIBentryALTinterwordspacing}{\spaceskip=\fontdimen2\font plus
\BIBentryALTinterwordstretchfactor\fontdimen3\font minus \fontdimen4\font\relax}
\providecommand{\BIBforeignlanguage}[2]{{%
\expandafter\ifx\csname l@#1\endcsname\relax
\typeout{** WARNING: IEEEtran.bst: No hyphenation pattern has been}%
\typeout{** loaded for the language `#1'. Using the pattern for}%
\typeout{** the default language instead.}%
\else
\language=\csname l@#1\endcsname
\fi
#2}}
\providecommand{\BIBdecl}{\relax}
\BIBdecl

\bibitem{delcroix2018single}
M.~Delcroix, K.~\v{Z}mol\'{i}kov\'{a}, K.~Kinoshita, A.~Ogawa, and T.~Nakatani, ``Single channel target speaker extraction and recognition with speaker beam,'' in \emph{Proc. IEEE International Conference on Acoustics, Speech and Signal Processing}, 2018, pp. 5554--5558.

\bibitem{wang2018deep}
J.~Wang, J.~Chen, D.~Su, L.~Chen, M.~Yu, Y.~Qian, and D.~Yu, ``Deep extractor network for target speaker recovery from single channel speech mixtures,'' in \emph{Proc. Annual Conference of the International Speech Communication Association}, 2018, pp. 307--311.

\bibitem{vzmolikova2019speakerbeam}
K.~{\v{Z}}mol{\'\i}kov{\'a}, M.~Delcroix, K.~Kinoshita, T.~Ochiai, T.~Nakatani, L.~Burget, and J.~{\v{C}}ernock{\`y}, ``{SpeakerBeam}: Speaker aware neural network for target speaker extraction in speech mixtures,'' \emph{IEEE Journal of Selected Topics in Signal Processing}, vol.~13, no.~4, pp. 800--814, 2019.

\bibitem{wang2018voicefilter}
Q.~Wang, H.~Muckenhirn, K.~Wilson, P.~Sridhar, Z.~Wu, J.~R. Hershey, R.~A. Saurous, R.~J. Weiss, Y.~Jia, and I.~L. Moreno, ``{VoiceFilter}: Targeted voice separation by speaker-conditioned spectrogram masking,'' in \emph{Proc. Annual Conference of the International Speech Communication Association}, 2019, pp. 2728--2732.

\bibitem{10113382}
K.~Zmolikova, M.~Delcroix, T.~Ochiai, K.~Kinoshita, J.~Černocký, and D.~Yu, ``Neural target speech extraction: An overview,'' \emph{IEEE Signal Processing Magazine}, vol.~40, no.~3, pp. 8--29, 2023.

\bibitem{wangvoicefilter}
Q.~Wang, I.~L. Moreno, M.~Saglam, K.~Wilson, A.~Chiao, R.~Liu, Y.~He, W.~Li, J.~Pelecanos, M.~Nika, and A.~Gruenstein, ``{VoiceFilter-Lite}: Streaming targeted voice separation for on-device speech recognition,'' in \emph{Proc. Annual Conference of the International Speech Communication Association}, 2020, pp. 2677--2681.

\bibitem{eskimez2022personalized}
S.~E. Eskimez, T.~Yoshioka, H.~Wang, X.~Wang, Z.~Chen, and X.~Huang, ``Personalized speech enhancement: New models and comprehensive evaluation,'' in \emph{Proc. IEEE International Conference on Acoustics, Speech and Signal Processing}, 2022, pp. 356--360.

\bibitem{thakker2022fast}
M.~Thakker, S.~E. Eskimez, T.~Yoshioka, and H.~Wang, ``Fast real-time personalized speech enhancement: End-to-end enhancement network ({E3Net}) and knowledge distillation,'' in \emph{Proc. Annual Conference of the International Speech Communication Association}, 2022, pp. 991--995.

\bibitem{dubey2022icassp}
H.~Dubey, V.~Gopal, R.~Cutler, A.~Aazami, S.~Matusevych, S.~Braun, S.~E. Eskimez, M.~Thakker, T.~Yoshioka, H.~Gamper, and R.~Aichner, ``{ICASSP} 2022 deep noise suppression challenge,'' in \emph{Proc. IEEE International Conference on Acoustics, Speech and Signal Processing}, 2022, pp. 9271--9275.

\bibitem{luo2019conv}
Y.~Luo and N.~Mesgarani, ``{Conv-TasNet}: Surpassing ideal time--frequency magnitude masking for speech separation,'' \emph{IEEE/ACM Transactions on Audio, Speech, and Language Processing}, vol.~27, no.~8, pp. 1256--1266, 2019.

\bibitem{liu2023quantitative}
X.~Liu, X.~Li, and J.~Serr{\`a}, ``Quantitative evidence on overlooked aspects of enrollment speaker embeddings for target speaker separation,'' in \emph{Proc. IEEE International Conference on Acoustics, Speech and Signal Processing}.\hskip 1em plus 0.5em minus 0.4em\relax IEEE, 2023, pp. 1--5.

\bibitem{gu2020hippo}
A.~Gu, T.~Dao, S.~Ermon, A.~Rudra, and C.~R{\'e}, ``{HiPPO}: Recurrent memory with optimal polynomial projections,'' in \emph{Proc. Advances in neural information processing systems}, vol.~33, 2020, pp. 1474--1487.

\bibitem{gu2021combining}
A.~Gu, I.~Johnson, K.~Goel, K.~Saab, T.~Dao, A.~Rudra, and C.~R{\'e}, ``Combining recurrent, convolutional, and continuous-time models with linear state space layers,'' in \emph{Proc. Advances in neural information processing systems}, vol.~34, 2021, pp. 572--585.

\bibitem{gupta2022diagonal}
A.~Gupta, A.~Gu, and J.~Berant, ``Diagonal state spaces are as effective as structured state spaces,'' in \emph{Proc. Advances in Neural Information Processing Systems}, vol.~35, 2022, pp. 22\,982--22\,994.

\bibitem{gu2021efficiently}
A.~Gu, K.~Goel, and C.~R{\'e}, ``Efficiently modeling long sequences with structured state spaces,'' in \emph{Proc. International Conference on Learning Representations}, 2022.

\bibitem{gu2022parameterization}
A.~Gu, K.~Goel, A.~Gupta, and C.~R{\'e}, ``On the parameterization and initialization of diagonal state space models,'' in \emph{Proc. Advances in Neural Information Processing Systems}, vol.~35, 2022, pp. 35\,971--35\,983.

\bibitem{nguyen2022s4nd}
E.~Nguyen, K.~Goel, A.~Gu, G.~Downs, P.~Shah, T.~Dao, S.~Baccus, and C.~R{\'e}, ``{S4ND}: Modeling images and videos as multidimensional signals with state spaces,'' in \emph{Proc. Advances in neural information processing systems}, vol.~35, 2022, pp. 2846--2861.

\bibitem{miyazaki2023structured}
K.~Miyazaki, M.~Murata, and T.~Koriyama, ``Structured state space decoder for speech recognition and synthesis,'' in \emph{Proc. IEEE International Conference on Acoustics, Speech and Signal Processing}, 2023, pp. 1--5.

\bibitem{shan2023augmenting}
H.~Shan, A.~Gu, Z.~Meng, W.~Wang, K.~Choromanski, and T.~Sainath, ``Augmenting conformers with structured state space models for online speech recognition,'' in \emph{Proc. IEEE International Conference on Acoustics, Speech and Signal Processing}, 2024 (to appear).

\bibitem{goel2022s}
K.~Goel, A.~Gu, C.~Donahue, and C.~R{\'e}, ``It’s raw! audio generation with state-space models,'' in \emph{International Conference on Machine Learning}, 2022, pp. 7616--7633.

\bibitem{ku2023multi}
P.-J. Ku, C.-H.~H. Yang, S.~Siniscalchi, and C.-H. Lee, ``{A Multi-dimensional Deep Structured State Space Approach to Speech Enhancement Using Small-footprint Models},'' in \emph{Proc. Annual Conference of the International Speech Communication Association}, 2023, pp. 2453--2457.

\bibitem{delcroix2020improving}
M.~Delcroix, T.~Ochiai, K.~Zmolikova, K.~Kinoshita, N.~Tawara, T.~Nakatani, and S.~Araki, ``Improving speaker discrimination of target speech extraction with time-domain {SpeakerBeam},'' in \emph{Proc. IEEE International Conference on Acoustics, Speech and Signal Processing}, 2020, pp. 691--695.

\bibitem{panayotov2015librispeech}
V.~Panayotov, G.~Chen, D.~Povey, and S.~Khudanpur, ``{LibriSpeech}: An {ASR} corpus based on public domain audio books,'' in \emph{Proc. IEEE International Conference on Acoustics, Speech and Signal Processing}, 2015, pp. 5206--5210.

\bibitem{reddy2021interspeech}
C.~K. Reddy, H.~Dubey, K.~Koishida, A.~Nair, V.~Gopal, R.~Cutler, S.~Braun, H.~Gamper, R.~Aichner, and S.~Srinivasan, ``{INTERSPEECH} 2021 deep noise suppression challenge,'' in \emph{Proc. Annual Conference of the International Speech Communication Association}, 2021, pp. 2796--2800.

\bibitem{tasnet}
``{conv-tasnet},'' \url{https://github.com/funcwj/conv-tasnet}, Cited February 28 2024.

\bibitem{statespace}
``{state-spaces},'' \url{https://github.com/state-spaces/s4}, Cited February 28 2024.

\bibitem{DCCRN}
``{DeepComplexCRN},'' \url{https://github.com/huyanxin/DeepComplexCRN}, Cited February 28 2024.

\bibitem{nagrani17_interspeech}
A.~Nagrani, J.~S. Chung, and A.~Zisserman, ``{VoxCeleb: A Large-Scale Speaker Identification Dataset},'' in \emph{Proc. Annual Conference of the International Speech Communication Association}, 2017, pp. 2616--2620.

\bibitem{chung18b_interspeech}
J.~S. Chung, A.~Nagrani, and A.~Zisserman, ``{VoxCeleb2}: Deep speaker recognition,'' in \emph{Proc. Annual Conference of the International Speech Communication Association}, 2018, pp. 1086--1090.

\bibitem{Resemblyzer}
``{Resemblyzer},'' \url{https://github.com/resemble-ai/Resemblyzer/tree/master}, Cited February 28 2024.

\bibitem{kingma2014adam}
D.~P. Kingma and J.~Ba, ``Adam: A method for stochastic optimization,'' in \emph{Proc. International Conference on Learning Representations}, 2015.

\bibitem{vincent2006performance}
E.~Vincent, R.~Gribonval, and C.~F{\'e}votte, ``Performance measurement in blind audio source separation,'' \emph{IEEE Transactions on Audio, Speech, and Language Processing}, vol.~14, no.~4, pp. 1462--1469, 2006.

\bibitem{reddy2022dnsmos}
C.~K. Reddy, V.~Gopal, and R.~Cutler, ``Dnsmos p. 835: A non-intrusive perceptual objective speech quality metric to evaluate noise suppressors,'' in \emph{Proc. IEEE International Conference on Acoustics, Speech and Signal Processing}, 2022, pp. 886--890.

\end{thebibliography}

\end{document}